# Why do nanowires grow with their c-axis vertically-aligned in the absence of epitaxy?

Almog R. Azulay,[1] Yury Turkulets,[1] Davide Del Gaudio,[2] Rachel S. Goldman,[2] and Ilan Shalish[1]*

[1]Ben Gurion University of the Negev, Beer Sheva, Israel    [2]University of Michigan at Ann Arbor

Images of uniform and upright nanowires are fascinating, but often, they are quite puzzling, when epitaxial templating from the substrate is clearly absent. Here, we reveal the physics underlying one such hidden growth guidance mechanism through a specific example - the case of ZnO nanowires grown on silicon oxide and glass. We show how electric fields exerted by the insulating substrate may be manipulated through the surface charge to define the orientation and polarity of the nanowires. Surface charge is ubiquitous on the surfaces of semiconductors and insulators, and as a result, substrate electric fields need always be considered. Our results suggest a new concept, according to which the growth of wurtzite semiconductors may often be described as a process of electric-charge-induced self assembly, wherein the internal built-in field in the polar material tends to align in parallel to an external field exerted by the substrate to minimize the interfacial energy of the system.

Nanowires often adopt specific orientation during their growth despite the lack of eptitaxial guidance from the substrate. While this phenomenon may not be limited to nanowires, it has become more easily apparent with nanowires, with most of the evidence coming from ZnO. Several unique features make ZnO one of the most intensively studied semiconductors today.[1] To date, ZnO has been used as a transparent conductor in photovoltaic applications,[2] varistor for voltage surge protection,[3] solar blind photodetector,[4] gas sensor,[5] and photocatalyst material.[6] It has also been proposed for several future applications, as transparent field effect transistors,[7] UV light-emitting diodes,[8] memristors,[9] biosensors,[10] and spintronic devices.[11,12] Growth of ZnO on Si is vital for integrating this material into the present microelectronic technology. One peculiar observation that has become more easily apparent with the growth of ZnO nanowires is that on substrates such as glass and silicon, that do not provide an epitaxial template, ZnO grows in a direction along the c axis, i.e. it is preferentially c-oriented.[13] For example, in **Fig. 1a**, the presence of 3 ZnO hexagonally-shaped mesas on a Si(100) substrate confirms that the mesas are indeed c-oriented. The sides of the hexagonal mesas are not mutually parallel, as shown schematically in **Fig. 1b**, thus confirming the lack of epitaxial relations with the substrate. Growth of c-oriented nanowires on Si substrate, as shown in **Fig. 1c**, has been reported in several papers. For example, c-oriented ZnO nanowire arrays have been grown on silicon or glass substrates without the use of a preexisting textured thin film. These were synthesized at temperatures ranging from 400 to 600 °C using metal organic vapor phase epitaxy,[14] chemical vapor deposition (CVD),[15] atomic layer deposition (ALD),[16] chemical vapor transport (CVT),[17] or hydrothermal approach.[18,19] In many cases, the upright alignment was explained by the formation of a ZnO wetting layer template for ZnO nanowire alignment; however, an explanation for the c-axis orientation of the wetting layer was not given.[17] Similarly, in Ref. 19 zinc acetate was spin-coated on the growth substrate. In that case, it was suggested that thermal decomposition of the zinc acetate produced ZnO colloids and nanocrystals with their (0002) planes parallel to the substrate surface. However, the question why these nuclei prefer a specific orientation has remained open.

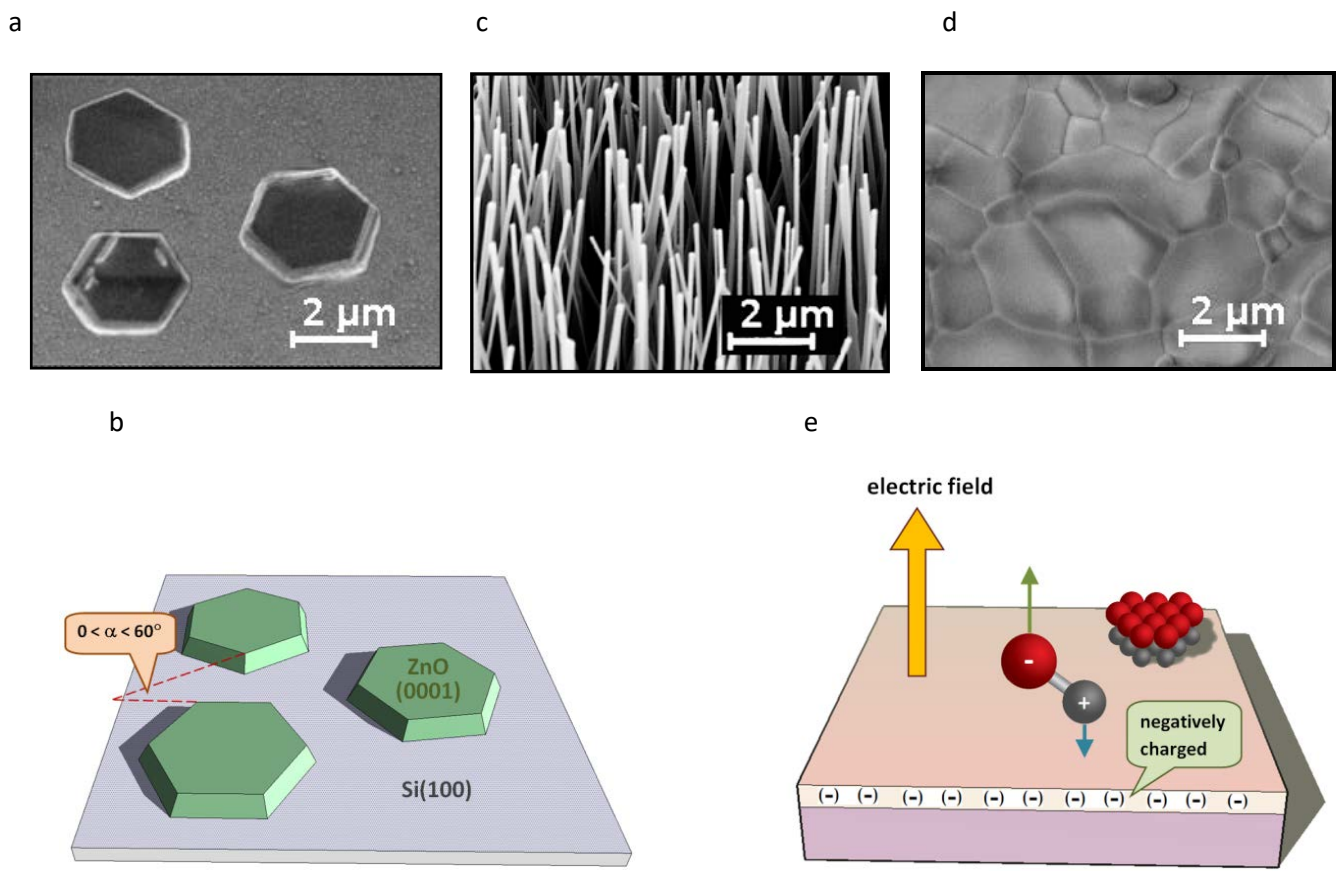

**Figure 1** – (**a**) SEM image acquired at sample-tilt of 30° of hexagonally shaped mesas grown on Si(100) substrate, and (**b**) schematic cartoon showing the rotational angle difference, which confirms the lack of epitaxial relations with the substrate, despite the alignment of the c-axis perpendicular to the substrate. (**c**) SEM image of c-oriented ZnO nanowires grown on Si(100) substrate (see Supplementary Information for angle statistics). (**d**) Polycrystalline ZnO film grown on a Si(100) substrate. Each individual grain is c-aligned but is rotated on the c-axis in a random angle. When they expand, they meet each other forming grain boundaries (see Supplementary Information for x-ray diffraction). (**e**) Schematic of Si substrate with negatively charged surface oxide, which exerts an electric field that aligns the polar axis of the deposited molecule along electric field lines. For ZnO, the zinc (oxygen) face is positively (negatively) charged; thus the oxygen face points up..

In contrast to the reported experimental work, several theoretical papers have addressed the problem of the observed oriented growth along the c-axis in attempts to reconcile it with the expected instability of the polar faces.[20,21,22,23,24] Wander et al. proposed that a near surface geometric relaxation, taking place gradually toward the surface over a depth of the several outmost layers along with charge transfer, stabilize the free polar surfaces of a bulk crystal.[20] This mechanism appears to require several crystal layers to take effect. Claeyssens et al. suggested a graphitic structure reconstruction of the ZnO of films thinner than 18 monolayers, over which c-oriented growth should be favored.[21] This reconstruction was required in order to cancel the polarization charge present on the polar faces to allow the ZnO to grow on a substrate that was not charged. It may well be that growth of a polar crystal on a non-polar, uncharged, substrate induces a reconstruction that removes the polar nature of the single monolayer altogether.

However, as our present results seem to suggest, such reconstruction may be altogether redundant in the special case of a charged substrate. As a matter of fact, in all of these theoretical studies, no specific substrate influence was clearly considered as part of the model. However, from the thermodynamic point of view, the substrate and the deposited crystals form an interface, and the minimization of the energy of this interface provides an important driving force. If the kinetics is not too fast and the system is allowed enough time to reach equilibrium, this driving force may become important enough to derive the arrangement of the deposited crystals on the surface. Here, we propose a simple mechanism that resolves the instability problem of the polar surfaces and renders the reconstruction redundant, at least when the substrate is Si, though the same effect may not be limited to ZnO on Si.

During vapor growth of ZnO, oxygen is an important part of the ambient. At the typically high growth temperatures, the Si substrate will oxidize faster than Zn, due to the lower free energy of formation of SiO2 (~-727 KJ/mole at 1000 °K) compared with that of ZnO (~-256 KJ/mole at 1000 °K).[25] As a result, SiO2 will form immediately at the very early stages of the growth process.

The presence of oxygen at the high growth temperature is also known to induce charging in the oxide layer. Charging of the oxide during its growth has been reported already in the early days of the field effect transistor[26] and is a known by-product of the oxide growth process.[27,28,29] Typically, a positive charge is induced during CVD or thermal growth, but may turn negative in plasma enhanced CVD.[30]

As schematically depicted in **Fig. 1e**, this surface charge induces an electric field perpendicular to the substrate surface. Since the polar charge on the oxygen face of ZnO is negative, the presence of an electric field should align the growth perpendicular to the substrate. Hence, a c-oriented growth is a reasonable outcome.

Since there are no epitaxial relations with the substrate, different individual nuclei may all have their c-axis perpendicular to the substrate but otherwise be randomly rotated. As the nuclei expand laterally, they meet one another forming a c-oriented polycrystalline film, as in **Fig. 1d**. Nanowire growth on such a substrate will result in upright nanowires. However, nanowires grown on different grains will have a variety of in-plane rotation angles, reflecting that of the

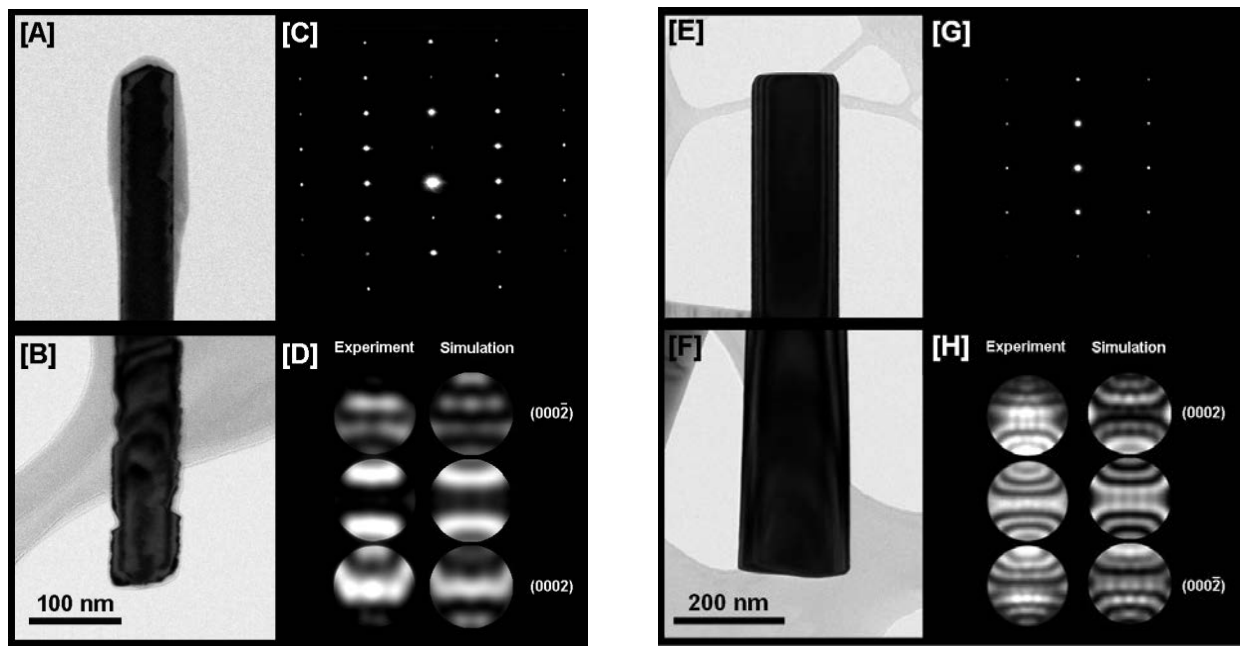

Figure 2 – **left** - [A] TEM of the tip-end of a single CVD-grown ZnO nanowire on thermal oxide (the sample of Fig. 2). [B] TEM image of the bottom/substrate end of the same wire. Note that the bottom of the wire is typically slightly wider with rougher surface. [C] Selective area electron diffraction (SAED), and [D] Experimental and simulated converging beam electron diffraction (CBED) corresponding to the diffraction in [C] acquired from the bottom end of Fig. B. The simulation corresponds to 70-nm-thick ZnO. **Right** - [E] TEM of the tip-end of a single CVD-grown ZnO nanowire on PECVD-grown oxide (the sample of Fig. 2). [F] TEM image of the bottom/substrate end of the same wire. [G] Selective area electron diffraction (SAED) shows that the wire is oriented along the c-axis, and [H] Experimental and simulated converging beam electron diffraction (CBED) corresponding to the diffraction in [G] acquired from the bottom end of Fig. F. The simulation corresponds to 170-nm-thick ZnO (the diffraction was taken slightly off the center of the wire). (see Supplementary Information for additional data)

"seed" grain.

For these studies, we utilized high-resistivity (~2000 Ohm·cm) boron-doped single-side polished Si(100) substrates. Charge in the surface oxide was verified using a Kelvin probe.[31] To form ZnO seed layers, Zn-acetate was deposited and annealed at 1020 °C for 30 min. Due to the vanishingly small quantity of ZnO produced by the seeding phase, attempts to characterize it with TEM where not successful, consistent with earlier reports. The samples were placed in a quartz crucible containing a 1:1 ZnO powder: graphite mixture and introduced into the pre-heated tube furnace using a linear-motion feed-through. The ZnO nanowires were then grown for 5 minutes at at ~1020 °C in Ar/O2 (30/4 sccm) ambient. TEM imaging and electron diffraction were carried out in a JEOL JEM-2100F instrument operating at 200 kV. CBED simulation was carried out using JEMS electron microscopy simulation software.[32] For each oxide type, growth was repeated 3 times. For each growth, TEM/CBED imaging was carried out on ~10 nanowires.

To test our hypothesis that c-oriented ZnO growth on SiO2 is caused by the oxide electric charge, we consider the sign of the charge. Specifically, we consider that control of the charge sign may cause (0002) growth in the case of negative charge (electron trapping), in contrast to the commonly observed (000-2) growth on the typical hole-trapping (positively charged) thermally-grown SiO2. Therefore, we grew ZnO nanowires on thermally-grown, and on plasma-enhanced CVD grown, SiO2. The presence and type of surface charge was verified using a Kelvin probe (see Supplementary Information).

**Figures 2a** and **2b** show transmission electron microscopy (TEM) images of the two ends of a CVD-grown ZnO nanowire grown on a positively charged thermal oxide. Diffractions were acquired at both ends of each wire to verify that no polarity inversion has taken place during the growth. The shown diffractions were obtained from the bottom/substrate end of the wire shown in **Fig 2b**. The bottom part of the wire shown in **Fig. 2b** is slightly thicker than the other end of the wire – a typical and common observation in nanowires that is useful for identification of the bottom end, especially for self-catalyzed nanowires which often lack a catalyst ball at their tip. **Figure 2c** shows the selective-area electron diffraction (SAED) pattern acquired from the bottom end of the wire, and **Fig. 2d** shows the corresponding convergent-beam electron diffraction (CBED) pattern side by side with the simulated CBED pattern revealing a (000-2)-oriented growth direction confirming the prediction of our model, i.e., on a negatively charged oxide, the Zn-terminated face (having negative polar charge) is the growth face, while the O-terminated face (positive polar charge) faces the substrate.

The right hand panel of **Fig. 2** (**Figs. 2e** through **2h**) follows the same structure as in the left hand panel of **Fig. 2** (**Figs. 2a** through **2d**) showing TEM and CBED results for a CVD-grown ZnO nanowire grown on a negatively charged PECVD oxide. Comparison of the CBED pattern with simulated pattern reveals a (0002)-oriented growth direction confirming the prediction of our model for this case as well.

These diffraction results show that our ZnO wires grew c-oriented in both cases but flipped over their polarity when the sign of the substrate charge was changed. The wire polarity aligns parallel to the direction of the electric field. This suggests that the $SiO_2$ substrate charge plays a pivotal role in directing the growth of the ZnO wires. Since electrical charging is common in insulators, and since built-in polar fields are present in many polar materials, the mechanism observed here may not be limited to the $SiO_2$\ZnO system. Charge induced during the growth of the insulating oxide substrate exerts a normal-to-substrate electric field that works to align the polar axis of the nucleating seed layer or nanowire material normal to the substrate. The polarity of the polar axis reverses with the sign of the substrate charge. In the case presented here, the polarity reversal and the normal-to-substrate alignment appear to be two manifestations of the same phenomenon. This mechanism appears to work as an unintentional, naturally occurring, electrophoresis on the deposited polar material. Due to their built in electric field and dipolar nature, wurtzite materials possess an inner mechanism that affects their self-assembly into crystals.[33]

Several studies on thin films of polycrystalline Cu showed preferred crystallographic orientation (i.e. crystallographic texture) when grown on amorphous $SiO_2$ substrates, although epitaxy was not possible.[34,35] In these cases, Cu, which has a face-centered-cubic structure, tends to prefer the (111) orientation. Annealing of these films close to their melting points often increases the (111) texture. This specific orientation of the copper on the $SiO_2$ surface appears to minimize the surface free energy. In general, the surface free energy $\bar{f}$ is given by[36]

$$\bar{\bar{f}} = \gamma + \sum_i \mu_i \Gamma_i \quad (1)$$

where $\gamma$ is the surface tension, $\mu_i$ is the chemical potential of component $i$, and $\Gamma_i$ is the adsorption of component $i$. Since, in the Cu case, the grown crystal is a metal, one may safely ignore any possible influence of an electric field induced by electric charges in the substrate and consider the differences in the surface chemical potential alone. However, for polar materials, the effect of an electric field cannot be ignored, and the added electrical potential component is expressed in a replacement of the chemical potentials, $\mu_i$, by electrochemical potentials, $\tilde{\mu}_i$, given by the Guggenheim relationship: [36,37]

$$\tilde{\mu}_i = \mu_i + q z_i \varphi_i \quad (2)$$

where $q$ is the elementary charge, $z_i$ is the ion charge number, and $\varphi_i$ is the electrical potential. The contribution of the electrical potential to the interfacial energy depends both on the electric field emanating from the substrate and on the built-in polar field in the unit cell of the polar material. Hence, the strength of this effect may depend on the strength of the built-in polar electric field.

Our experiment shows a clear correlation of the ZnO growth polarity with the sign of the substrate charge. Previous studies have suggested that the chemical potential on the Zn-face of ZnO is quite different from that on the O-face.[38] This would imply that in the absence of electrical potentials, the ZnO system would prefer one growth polarity over the other, due to this difference in the chemical potentials. Indeed, growth of ZnO has been reported to prefer the (0002) polar orientation over (000-2).[39,40] In our case of an intentionally charged substrate, we observed growth in both polarities depending on the polarity on substrate charge. This suggests that the effect of the electrical potentials on the interface energy of the ZnO/SiO$_2$ system prevails over that of the chemical potentials. Thus, in spite of the difference in chemical potentials between the oxygen and the zinc-terminated polar faces of ZnO, the surface energy minimization achieved by the electrical alignment of the deposited polar molecules is apparently much greater than that which could be achieved by chemical potentials alone.

As previously discussed, several theoretical studies have considered the expected instability of the polar faces of ZnO and suggested that reconstruction in inevitable to explain their observed stability.[20,21,22,23,24] These studies have not considered the possible effect of specific substrates. Our results suggest that substrate charge may as well stabilize the polar face. Thus, the influence of the substrate electric field is likely to provide additional control over the growth of polar materials.

The substrate electric field applies torque to the ZnO dipole. The potential energy of the dipole is minimized only when the dipole is aligned parallel to the electric field. Since the electric field is aligned perpendicular to the substrate, the polar axis of the ZnO wires also aligns perpendicular to the substrate, resulting in the observed normal-to-surface growth.

To facilitate the polar alignment, it seems necessary to grow the nanowires in a two step process. In the first step, or the nucleation step, it is necessary to limit the amount of the nucleating material, to enable the formation of a seed layer, wherein the seeds are aligned. In the second step, epitaxial growth of nanowires may take place on the seed layer. If the first step is bypassed, the typically high nanowire growth rate disturbs the alignment. In other words, the kinetics may be too fast, and the system is not allowed enough time to reach equilibrium. In contrast, when using the Zn-acetate method, the quantity of Zn that is available for nucleation appears to be small enough to allow optimal alignment of the particles that may then serve as seeds for epitaxial growth of ZnO nanowires.

The observed correlation of polar direction of ZnO nanowires with the sign of the SiO$_2$ substrate charge suggests that the alignment of the ZnO polar axis perpendicular to the substrate is the effect of the electric field emanating from the substrate. The proposed mechanism may not be limited to the ZnO/Si system. It sets forth a new concept, according to which substrate charge may be exploited to affect the growth direction of polar semiconductors.

## Supplementary Information

Additional data is provided in the Supplementary Information.

## Acknowledgement

We gratefully acknowledge the support of BSF grant #2015700 and NSF grant # ECCS-1610362. We acknowledge the Ilse Katz Institute for Nanoscale Science & Technology at Ben Gurion

# Why do nanowires grow with their c-axis vertically-aligned in the absence of epitaxy?

Almog R. Azulay,[1] Yury Turkulets,[1] Davide Del Gaudio,[2] Rachel S. Goldman,[2] and Ilan Shalish[1]*

[1]Ben Gurion University of the Negev, Beer Sheva, Israel    [2]University of Michigan at Ann Arbor

## Supplementary Information

**Nanowire Angle**

Angles of 107 wires from Fig. 1c were measured using the software ImageJ.[1] The observed angles were 90.95±4.65 degrees. **Fig S1** shows a bar graph of the frequencies with 5 degree binning. Typically, nanowires will keep their vertical alignment at aspect ratios of about 10:1. At aspect ratios exceeding 100:1, they often bend by a few degrees, especially when grown on an insulator and imaged using electrons.

**X-ray Diffraction for Fig. 1d**

**Figure S2** shows two-theta x-ray diffraction obtained from the sample of Fig. 1(d) in the paper. This ZnO layer was grown on a thermally oxidized Si(100) with oxide thickness of about 2 µm. The diffraction shows a single peak at 34.38° which fits the (0002) diffraction in R050492 powder diffraction file.[2] This shows that albeit the polycrystalline appearance, the grains are c-oriented.

**Oxide charge**

In this work, Si wafers were thermally oxidized to produce a positively-charged oxide and PECVD-oxidized to produce negatively charged oxide. To verify the oxide charge, we conducted multiple measurements of the contact potential on 10 samples before oxidation, 10 samples after thermal oxide growth, and 10 samples after PECVD oxide growth. The Kelvin probe is brought to a distance of 0.3 mm from the sample surface forming a capacitor wherein one plate is the sample and the other plate is the probe. The probe is vibrated at a constant frequency and this varies sinusoidally the clearance between the plates (d). This variation of d produces an alternating (AC) current in the capacitor. The voltage on the probe is then changed until it matches the surface voltage on the sample, at which point the charge on the capacitor is zero and the AC current nullifies. This way the surface voltage (also known as the "contact potential") of the sample is measured. If a charged oxide layer is added within this capacitor, it will make the surface voltage more positive relative to its original value in the case of positively charged oxide, and more negative in the case of negatively charged oxide.

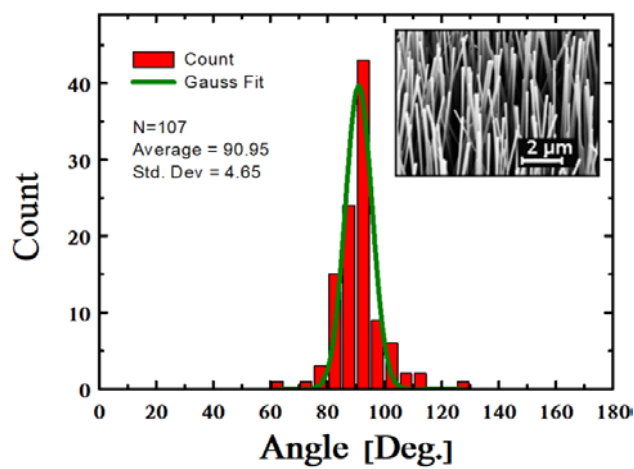

**FIG. S1** Histogram of angles of the nanowires in Fig. 1(c) in the paper. Angles were measured using the ImageJ software on N=107 nanowires. The average angle was 90.95 ± 4.65 °.

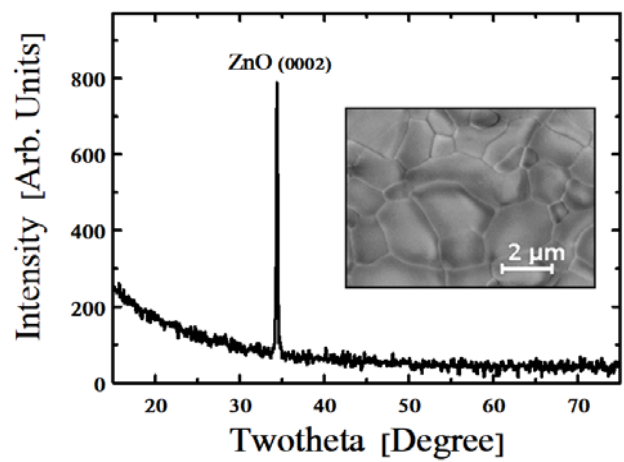

**FIG. S2** Two-theta x-ray diffraction from the Si(100)\SiO2\ZnO sample of Fig. 1(d) (SEM image of the surface is shown in the inset). The diffraction shows a single peak that fits the position of ZnO(0002) in R050492 powder diffraction file.

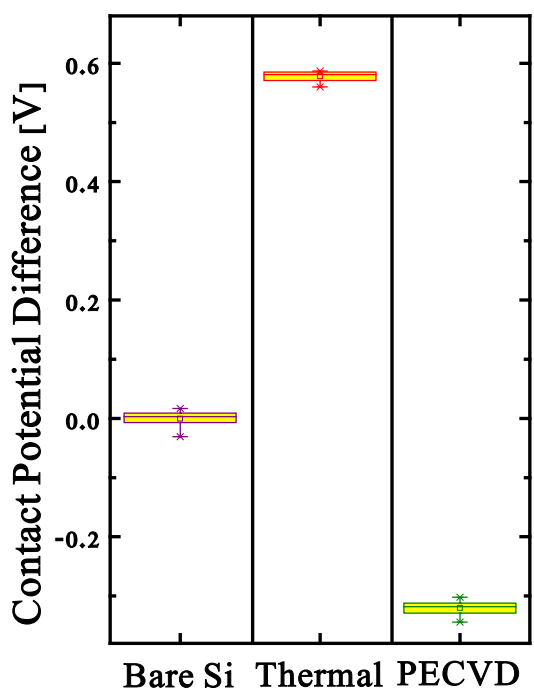

FIG. S3 Box plots of contact potential measured using a Kelvin probe on the oxidized Si wafers before ("Bare Si") and after thermal oxidation ("Thermal") or PECVD oxide growth ("PECVD"). The average value of contact potential before oxidation is set as zero for comparison.

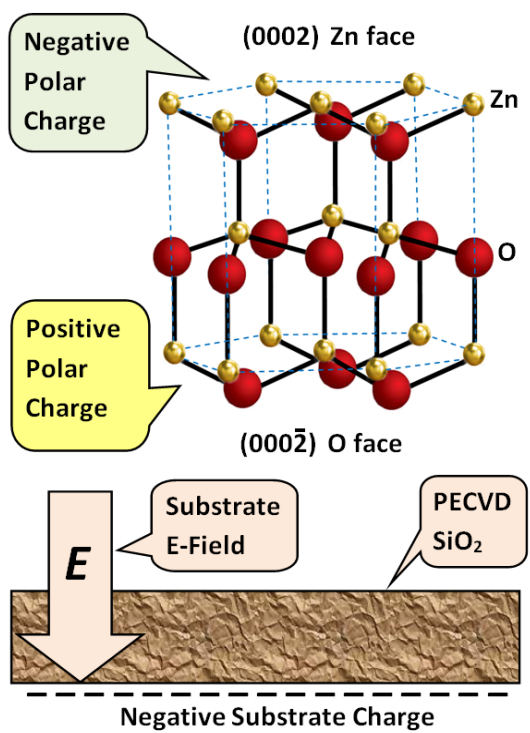

FIG. S4 Ball and stick model of the ZnO unit cell showing the polar crystal faces, their orientation, their polar charges, and the resulting growth orientation when grown of a negatively charged substrate.

**Figure S3** shows a statistical box plot for each of the statistical samples relative to the mean value measured before oxidation. The figure shows that thermal oxidation makes the contact potential more positive, while PECVD oxide makes it more negative. This confirms qualitatively the presence of positive charges in the thermal oxide and negative charges in the PECVD oxide, as expected. This result confirms in our samples what has been known for many decades since the early days of the transistor.[3,4,5,6] The theory underlying the Kelvin probe is described in great detail in a review by Kronik and Shapira.[7] Additional review of various oxide charge measurement methods using a Kelvin probe are described in a textbook by Schroder.[8]

**Growth Orientation vs. Substrate Charge**

In **Fig. S4**, we show a "ball and stick" structure of a ZnO unit cell. The Zn-terminated polar face is in the (0002) direction and is hosts negative polar charge, while the O-terminated face points in the (000-2) direction and hosts a positive polar charge. A polar crystal may be able to stabilize its positively charged face by placing it face-down on a negatively charged substrate. This explains why on PECVD-grown oxide (negatively charged), we observe the growth to be oriented in the (0002) direction (Zn-terminated face is up).

**Additional TEM/CBED results**

In Fig. S5, we add additional CBED data. All the data on each of the oxides, here and in the paper, were obtained from the same growth run. In thin wires the CBED pattern is blurred, and this makes it more difficult to compare it to simulation (see e.g. the thermal oxide CBED in line 4 below). For this reason, most of the CBED were acquired from thick wires. For each wire, CBED was acquired from both the bottom and the top of the wire to make sure that the polarity is identical throughout the wire. However for brevity, we show only a single pattern. In each figure, panel A shows the top end of the wire, panel B shows the bottom end, panel C shows the SAED pattern, and panel D compares the CBED pattern with a simulated CBED pattern.

**FIG. S5** Additional TEM/CBED results obtained from the same growth run as the results shown in Fig. 2 in the paper

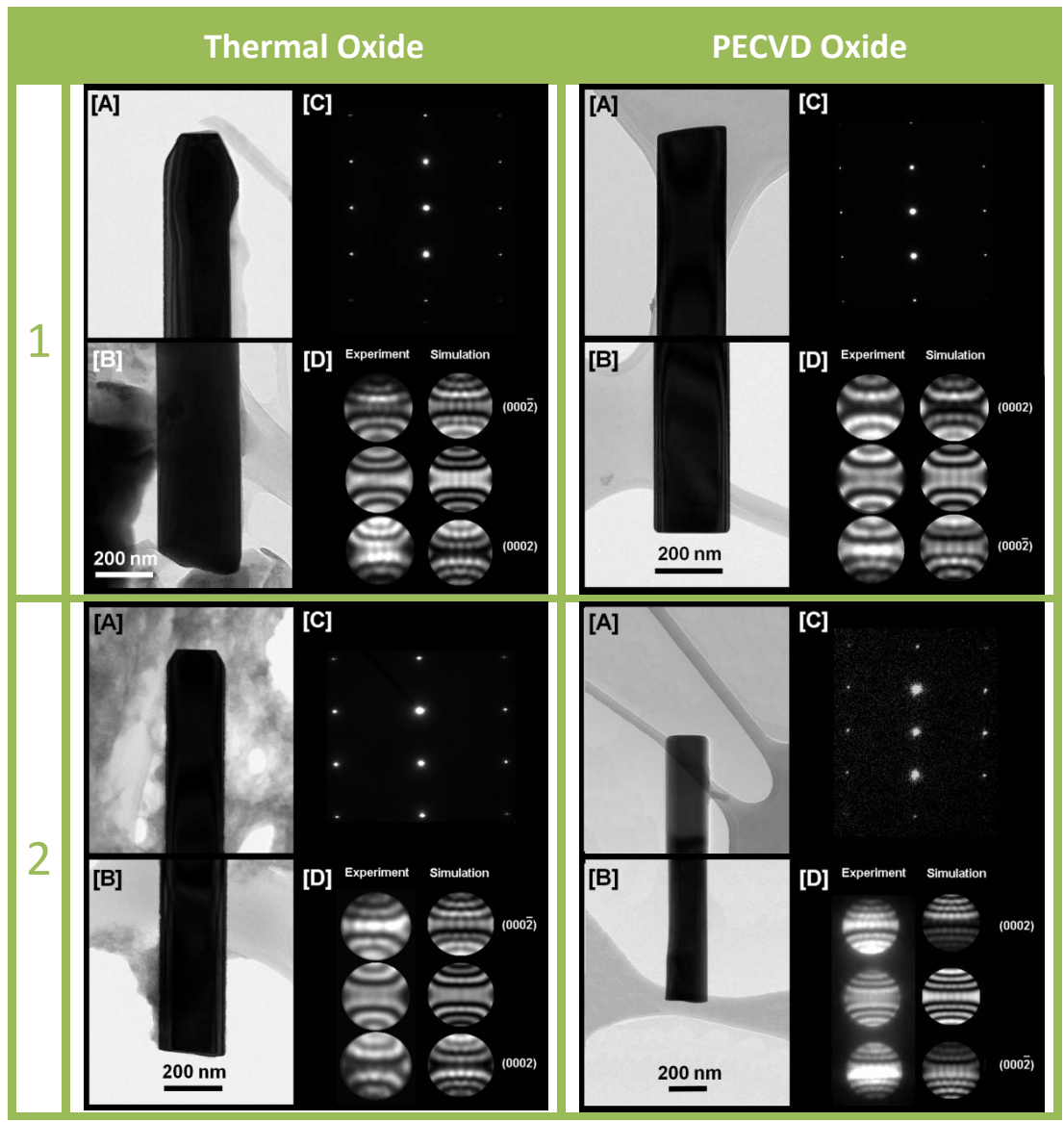

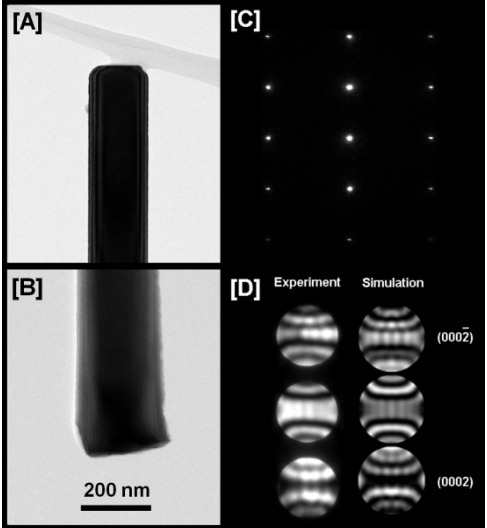
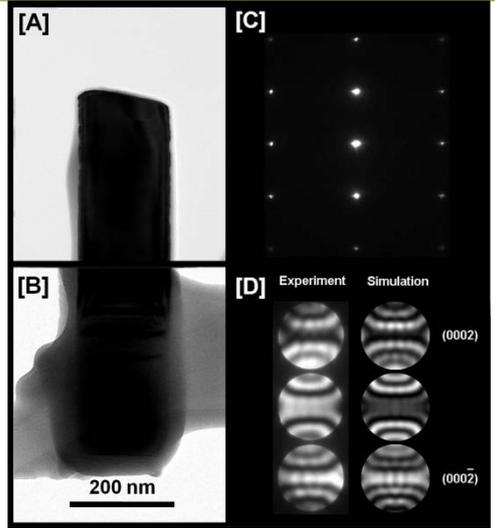
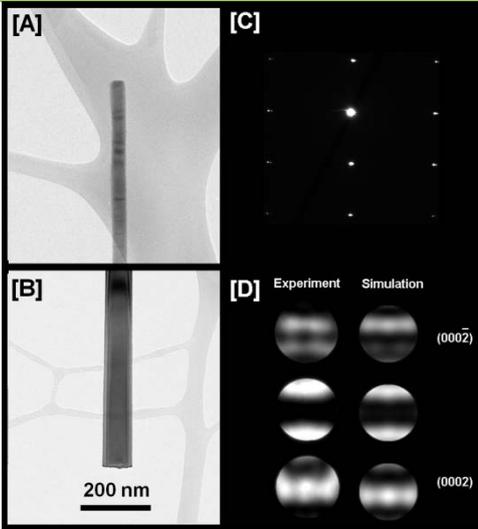
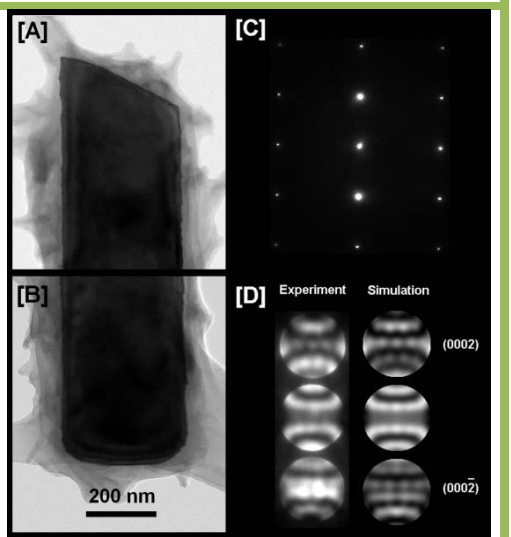
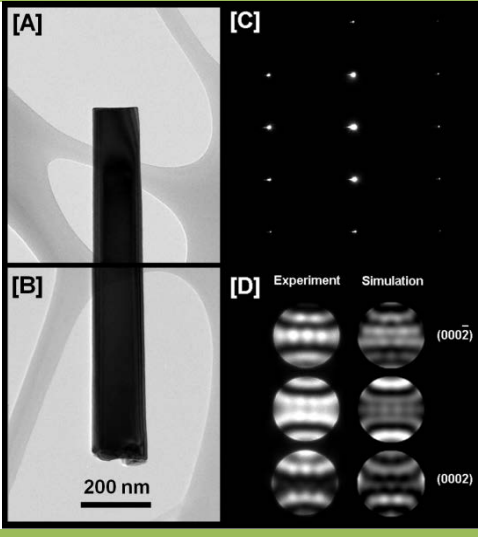
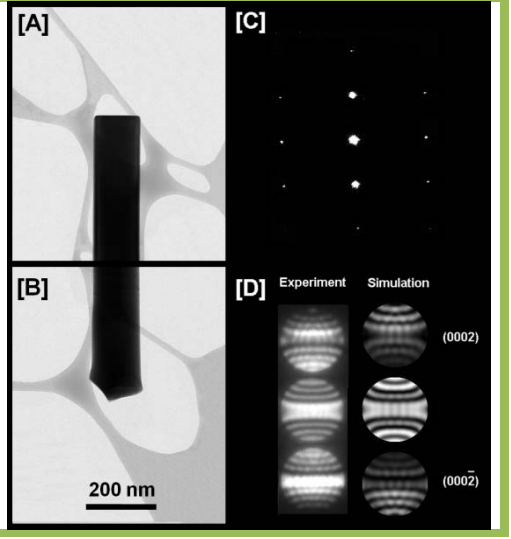